\documentclass[11pt,preprint]{aastex}
\usepackage{lscape}
\begin{document}

\title{{\sl Blood ties}: the real nature of the LMC binary globular clusters 
NGC~2136 and NGC~2137
\footnote{Based on observations collected at the ESO-VLT under the program 084.D-0933.}}

\author{Alessio Mucciarelli}
\affil{Dipartimento di Astronomia, Universit\`a 
degli Studi di Bologna, Via Ranzani, 1 - 40127
Bologna, ITALY}
\email{alessio.mucciarelli2@unibo.it}

\author{Livia Origlia}
\affil{INAF - Osservatorio Astronomico di Bologna, Via Ranzani, 1 - 40127
Bologna, ITALY}
\email{livia.origlia@oabo.inaf.it}

\author{ Francesco R. Ferraro}
\affil{Dipartimento di Astronomia, Universit\`a 
degli Studi di Bologna, Via Ranzani, 1 - 40127
Bologna, ITALY}
\email{francesco.ferraro3@unibo.it}

\author{Michele Bellazzini}
\affil{INAF - Osservatorio Astronomico di Bologna, Via Ranzani, 1 - 40127
Bologna, ITALY}
\email{michele.bellazzini@oabo.inaf.it}

\author{Barbara Lanzoni}
\affil{Dipartimento di Astronomia, Universit\`a 
degli Studi di Bologna, Via Ranzani, 1 - 40127
Bologna, ITALY}
\email{barbara.lanzoni3@unibo.it}

\begin{abstract} 
We have used a sample of high-resolution spectra obtained with the multi-fiber facility FLAMES
at the Very Large Telescope of the European Southern Observatory, to derive  
the kinematical and chemical properties of the two young Large Magellanic Cloud 
globular clusters NGC~2136 and NGC~2137. These two clusters represent a typical example of LMC
cluster pair suspected to be bound in a binary system: indeed the cluster centers of gravity 
have an angular separation  of less than 1.4 arcmin in the sky. 
The spectral analysis of seven giants in NGC~2136 and four in NGC~2137 reveals that the two clusters
share very similar systemic radial velocities, 
namely
$V_{r}$=~271.5$\pm$0.4 km/s ($\sigma$=~1.0 km/s) and
$V_{r}$=~270.6$\pm$0.5 km/s ($\sigma$=~0.9 km/s) for NGC~2136 and NGC~2137, respectively, 
and they have also indistinguishable abundance patterns.
The iron content is [Fe/H]=~-0.40$\pm$0.01 dex ($\sigma$=~0.03 dex) for NGC~2136
and -0.39$\pm$0.01 dex ($\sigma$=~0.01 dex) for NGC~2137, while the [$\alpha$/Fe] ratios are 
roughly solar in both clusters.
These findings suggest that the two clusters are gravitationally bound
and that they formed from the fragmentation of the same molecular cloud that was 
chemically homogeneous. This is the 
first firm confirmation of the binary nature  of a LMC cluster pair.
The most likely fate of this system is to merge into a single 
structure in a time-scale comparable with its orbital period.

\end{abstract}


\section{Introduction}

One of the most intriguing feature of the Large Magellanic Cloud (LMC) 
globular clusters (GCs) system is the large population of binary clusters 
that this galaxy harbors. The catalogue by \citet{dieball} includes 
a total of 473 candidate multiple (binary or triple) stellar clusters and associations 
with angular distances $\le$1.4 arcmin (corresponding to a projected distance of 20 pc 
when a distance modulus of 18.5 mag is assumed) and ages less than $\sim$1 Gyr. 
This sample corresponds to about 
10\% of the entire stellar clusters population in the LMC.
Similar binary systems are observed also in other galaxies, like the Small Magellanic Cloud (SMC) 
\citep{hatzi90}, M31 \citep{holland} and NGC~5128 \citep{minniti04}. 
In contrast, in our Galaxy the only recognized case is the open cluster pair NGC~869/NGC~884.

There are some possible scenarios to explain the nature of the binary clusters:
(i)~two clusters at different distances appear as a binary system only due to 
projection effects, lying along the same line of sight; 
(ii)~the clusters were born independently from distinct 
molecular clouds (likely with different ages and chemical compositions) and 
subsequently became a bound system after a close encounter or a tidal capture; 
(iii)~the clusters were born from the same molecular cloud (hence with the same age and metallicity) 
and are gravitationally bound.
Based on statistical arguments, \citet{dieball} suggest that the LMC binary clusters 
population cannot be simply explained in terms of apparent pairs, but a relevant 
fraction must be bound systems.

While the large number of binary clusters among the young Magellanic
globulars is a significant clue, the small projected distance on the sky 
between two clusters cannot guarantee the effective {\sl blood tie} of the objects.
Hints on binarity from the ages and photometric 
metallicities (with uncertainties at a level of 0.2 dex) have been provided 
for some tens of cluster pairs 
\citep[see for instance][]{dieball98,hilker,vallenari}. 
However,  
a firm validation of that binarity can only be obtained 
from the combined information of age, chemical 
abundances from high resolution spectra and radial velocities ($V_{r}$) 
measurements in order to clarify if the system is gravitationally bound
and unveil the possible common origin from the same molecular cloud.
At present such a binarity validation has not been probed in any 
LMC cluster pair.

This paper is devoted to address the true nature of the cluster pair
NGC~2136/NGC~2137. These are two young GCs with an angular separation of 
1.34 arcmin \citep{bhatia91}, corresponding to a projected separation of 
$\sim$19.5 pc, assuming a distance of 50 kpc (see left panel in Fig.~\ref{map}). 
Previous studies stated that the two clusters 
share the same age \citep[$\sim$80-100 Myr,see][]{hilker,dirsch} but 
no direct chemical and kinematical measurements are available to date.

\section{Observations}
Observations were performed with the multi-fiber facility FLAMES mounted at the 
ESO Very Large Telescope in the combined mode UVES+GIRAFFE. Data were acquired 
under a program devoted to investigate the chemical composition of the LMC GCs 
and their surrounding fields
The employed grating configuration 
includes the setups HR~11 and HR~13 for GIRAFFE and the 580 Red Arm 
for UVES. A total of 5 exposures of $\sim$45 min each was secured on the same 
targets configuration. 
The reduction of the spectra (including bias subtraction, flat-fielding, 
wavelength calibration and spectra extraction) was performed with the standard 
ESO pipelines.
Typical signal-to-noise ratio per pixel of the final coadded spectra is of $\sim$50-60 .
The FLAMES fibers were allocated on giant stars of the two clusters NGC~2136 and NGC~2137 
and of the surrounding field. 
Targets in the innermost 2.5 arcmin from the cluster center have been selected from the 
SofI near-infrared catalog by \citet{m06} while outermost objects were chosen from the 2MASS dataset.

\begin{figure}
\plotone{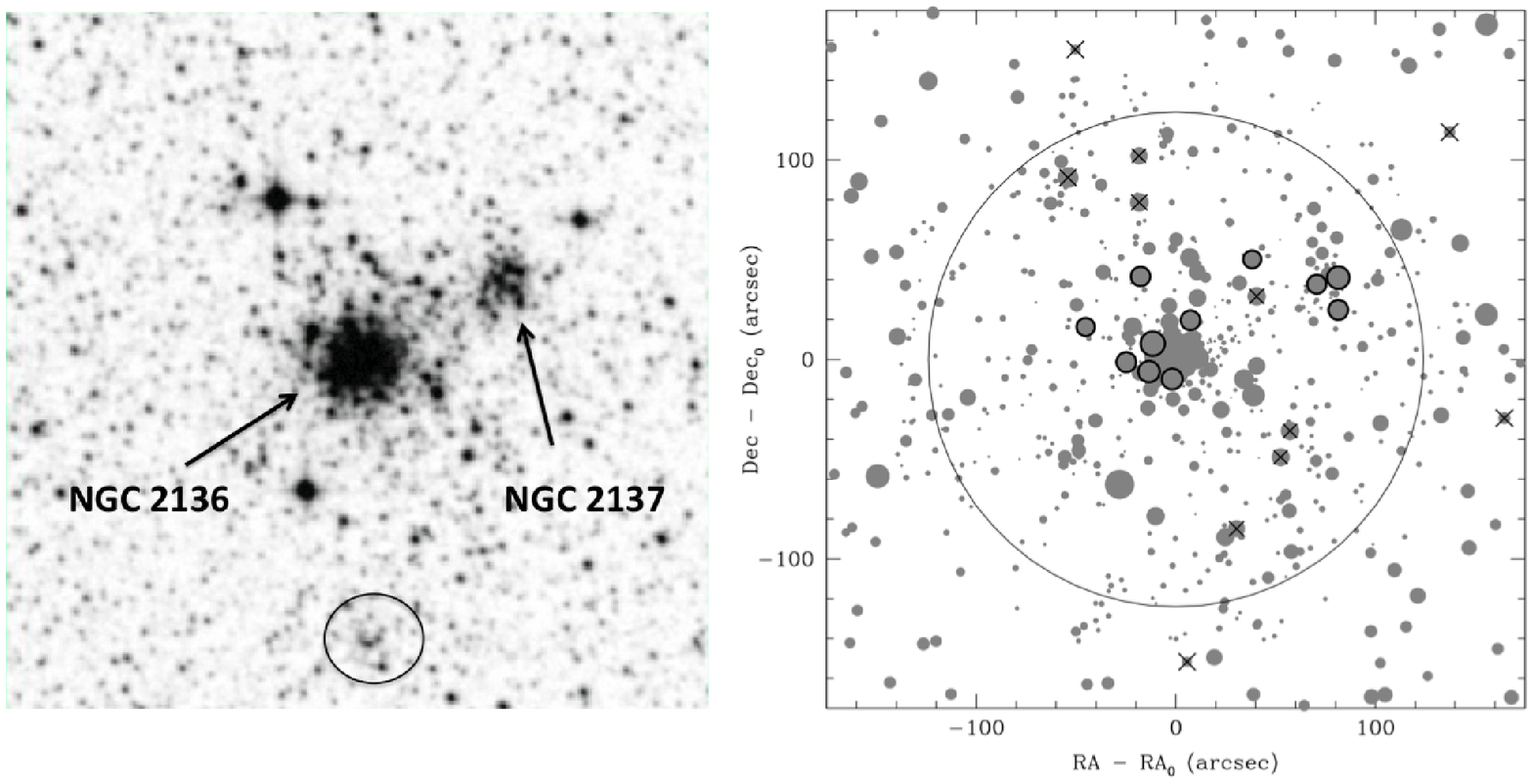}
\caption{{\sl Left panel}: V-band image of the NGC~2136/NGC~2137 system from the 
Digital Sky Survey archive. The circle marks the position of the possible
third component discussed by \citet{hilker}.
{\sl Right panel}: map of the IR catalog \citep[][ see text]{m06};
coordinates are referred to the
RA and Dec of the center of NGC~2136 \citep[see][]{m06}. The FLAMES targets discussed in 
this paper are marked as big grey circles. 
Crosses indicate observed GIRAFFE targets excluded because they belong to the Galaxy 
or to the LMC field. The large circle indicates the tidal radius by \citet{mclau}.}
\label{map}
\end{figure}

It is worth noting that the size of the clusters, their 
angular separation and the physical size of the FLAMES magnetic buttons 
do not allow to allocate a large number of fibers on the area covered by two clusters. 
A total of 22 fibers were finally allocated  in the inner 
3 arcmin (marked in the right panel of Fig.~\ref{map}). Here we discuss the kinematical 
and chemical properties of these stars.

Interestingly enough, \citet{hilker} suggest a possible third component of the system, identifying a faint 
stellar association located at a distance of $\sim$2.4 arcmin from the main cluster and 
embedded in a common stellar halo (the position of this stellar association is highlighted 
in the left panel of Fig.~\ref{map}). As apparent from the right panel of Fig.~\ref{map}, 
one FLAMES fiber was allocated also on this loose clump of stars.

\section{Radial Velocities and chemical analysis}

Radial velocities were derived by using the DAOSPEC code \citep{stetson} 
and measuring $\sim$370 and $\sim$140 absorption lines of different elements 
in the UVES and GIRAFFE spectra, respectively. Typical internal errors
(computed as $\sigma/\sqrt{N_{lines}}$) are of 0.05 for UVES and 0.15 km/s 
for GIRAFFE.
The accuracy of the zero-point of the wavelength calibration is checked 
by measuring several sky emission lines and compared with their restframe positions 
listed by \citet{ost}: the uncertainty in the zero-point 
has been added in quadrature to the internal error.

The chemical analysis was performed by using the suite of codes by R. L. Kurucz 
\citep[see][]{sbordone04}, aimed 
to compute model atmospheres, abundances from the observed equivalent widths and synthetic 
spectra. 
The employed model atmospheres were computed with ATLAS9, assuming plane-parallel geometry, 
LTE for all the species and no overshooting.

We used suitable linelists checked against spectral blending, in order to 
include only transitions predicted to be unblended for the corresponding 
spectral resolution and parameters. Oscillator strengths are from the most recent
version of the Kurucz/Castelli linelist 
\footnote{http://wwwuser.oat.ts.astro.it/castelli/linelists.html}. 
Equivalent widths were measured with DAOSPEC.
Atmospheric parameters were derived spectroscopically, 
by requiring {\sl (i)}~no trend between excitation potential and iron abundance 
to constrain the temperature; {\sl (ii)}~no trend between line strength and iron 
abundance to constrain the microturbulent velocity, and {\sl (iii)}~the same abundance 
from neutral and single ionized iron lines to constrain the gravity.

We measured abundances for Fe, Ni, Mg, O, Al, Na, Si, Ca, Ti. 
Oxygen abundances are derived through spectral synthesis 
of the forbidden line at 6300 $\mathring{A}$. Na abundances were derived from the doublet 
at 6154-60 $\mathring{A}$ for all the stars and also from the line at 5688 $\mathring{A}$ for 
the stars observed with UVES; departures from LTE were corrected following 
\citet{lind}. The total uncertainty for each abundance ratio was computed 
by adding in quadrature the internal error (computed as $\sigma/\sqrt{N_{lines}}$) 
and the uncertainty arising from the atmospheric parameters 
\citep[the latter computed following the approach by][]{cayrel04}.

\section{Results}

The heliocentric radial velocity $V_{r}$ distribution for all the 22 giants measured 
in the region of the two clusters
is shown in the upper panel of Figure 2. As can be seen, it is highly peaked at
$V_{r}\sim 270$ km/s. 
The lower panel of Figure 2 shows metallicities and radial velocities for the stars
with $V_{r}> 230$ km/s. 
A clear, well defined clump of 11 stars 
sharing virtually
the same radial velocity ($V_{r}\sim$ 270 km/s) and metallicity ($[Fe/H]\sim$--0.40) is visible.
Main information for these stars are listed in Table 1 and their position in the SofI color-magnitude 
diagram is shown in the left panel of Fig.~\ref{cmd}, together with portions of UVES spectra in 
the right panel. They are also marked in the right panel of Fig.~\ref{map}. The close inspection of this map allows
to attribute each star to one of the two clusters: 
we have attributed 7 stars to NGC~2136 and 4 stars
to NGC~2137\footnote{Note that star \#51 is located in the halfway between 
the two clusters and we decide to consider it member of NGC~2137.}.

Considering the entire sample of 11 stars,
the mean radial velocity   turns out to be 
of 271.2$\pm$0.3 km/s with a dispersion $\sigma$=~1.0 km/s. When we consider separately
the stars belonging to each individual cluster we obtain  $V_{r}=271.5\pm0.4$ km/s ($\sigma$=~1.0 km/s) 
 and 270.6$\pm$0.4 km/s ($\sigma$=~0.9 km/s) for NGC~2136 and NGC~2137, respectively. The two measurements
 are fully compatible at a level of 1.6$\sigma$\footnote{If 
 we attribute star \#51 to NGC~2136 (instead of NGC~2137) the results 
change by $\sim$0.1 km/s only.} 
and also compatible with the LMC velocity distribution, which peaks
at $\sim$257 km/s \citep[$\sigma$=~25 km/s,][]{cole05} .
No previous measurements of radial velocities of member stars in these clusters 
were available in the literature. 

The average iron content is [Fe/H]=~--0.40$\pm$0.01 dex and --0.39$\pm$0.01 dex for 
NGC~2136 and NGC~2137 respectively. These are the very first direct measurements of chemical abundances in
these clusters since
previous estimates of the metallicity   were based only on 
photometry. Both \citet{hilker} and \citet{dirsch} derived lower metallicities 
([Fe/H]=~-0.55 dex with a typical uncertainty of about 0.2 dex).
Table 1 lists the abundance ratios for each individual star and
Table 2 lists the average abundances for the measured chemical elements.
As can be seen the two clusters share virtually identical
abundance patterns either in terms of iron, $\alpha-$  and light elements.  In particular,
all the [$\alpha$/Fe] abundance ratios are roughly solar, 
pointing out that the gas from which both clusters formed has been enriched also by 
Type Ia Supernovae ejecta, as expected given their young age.

{\it This is the first clear-cut indication that stars in the two clusters are virtually 
indistinguishable both in terms of metallicity and radial velocity. Also, this is the first 
time that the real nature of a binary cluster is revealed.}

Note that the star observed in the possible third component of the 
system (plotted as a black circle in the left panel of Fig.~\ref{map}) has 
$V_{r}=281.4$ km/s and an iron content [Fe/H]=~--0.54 dex, incompatible with the chemical and
kinematical properties of the two clusters: this seems to exclude a link between this 
star and the two clusters.
At present, we are not able to assess if this stellar association is really bound to 
the other two clusters, hence we exclude it from the following discussion.

\section{Discussion}
 
From the analysis of the kinematical and chemical properties of the two clusters 
we obtained two relevant findings:

{\it (i) - All the measured stars share very similar $V_{r}$: indeed, the
 difference between the average values in the
two clusters is only $\Delta V_{r}= 0.9$ km/s.} 
Such a small value well agrees with the criterion by \citet{vdb} 
for the gravitational link of two stellar systems.
In order to compute the orbital velocity 
of the binary system, we first estimated the cluster masses.
The SIMBAD database provides integrated V magnitudes of 10.70 and 12.66 
for the main and secondary cluster respectively, corresponding to 
$L_{V}^{2136}=1.13$ $10^5 L_{\odot}$ and 
$L_{V}^{2137}=1.85$ $10^4 L_{\odot}$. 
Adopting the mass-to-light ratio of $M/L_{V}$=0.119, appropriate for a simple 
stellar population of 100 Myr,
Z=0.008 and solar-scaled chemical composition computed from 
the BaSTI database\footnote{http://albione.oa-teramo.inaf.it/},
we obtain masses of $M_{2136}= 1.34$ $10^4 M_{\odot}$ and 
$M_{2137}= 0.22$ $10^4 M_{\odot}$ (with a  mass ratio of 0.16)
\footnote{Note that 
\citet{mclau} derived a slightly higher mass for
NGC~2136 by fitting the
surface brightness profile: they obtained $M_{2136}$= 1.99 $10^4 M_{\odot}$, 
2.19 $10^4 M_{\odot}$
and  2.09 $10^4 M_{\odot}$ adopting the King, power-law and Wilson model, respectively.
Since no estimate was obtained for NGC~2137, in the following 
 we will adopt  the  masses derived from the integrated V magnitudes.}. 
With these mass values and by assuming a circular orbit and $R_{3D}=\sqrt{3/2}R_{p}$, 
(where $R_{p}$ is the projected distance, with $R_{p}$=~19.5 pc) as a statistical proxy 
of the de-projected,three-dimensional distance, the orbital period can be easily inferred 
from the third Kepler's law, yielding a value of $P_{orb}\sim$87 Myr. 
Finally, the orbital velocity (computed as $V_{orb}=2\pi R_{3D}/P_{orb}$) 
turns out to be $V_{orb}\sim$1.7 km/s .  
This value indicates the maximum expected difference between the velocities 
of the two clusters. 
Such a small value is in excellent agreement with the observations and 
enforce our statement that the two objects are gravitationally bound.

{\it (ii) - The two clusters share the same chemical abundances and abundance pattern, 
in terms of iron and $\alpha-$ elements, 
suggesting that 
they likely formed
from the collapse of the same (chemically homogeneous) molecular cloud}.
Interestingly enough, the two clusters also show similar and homogeneous abundances 
of the light elements (Na, O, Mg and Al), 
at variance with the old GCs in our Galaxy \citep[see e.g.][]{carretta09} and in the LMC 
\citep{m09}, which show clear spreads and some anti-correlations.
Such a lack of abundance spread among light elements has been already found    
in a few other LMC GCs of young/intermediate ages 
\citep{f06,m07,m08}
with metallicity similar to the NGC~2136/NGC~2137 pair. 
From this point of view, the LMC GCs younger than $\sim$3 Gyr behave like 
the Galactic open clusters, that do not show evidences of abundance anomalies
\citep[see e.g.][]{desilva,martell}.

This finding seems to indicate that while older globulars in both our Galaxy and in the LMC 
self-enriched at the very early stage of their formation 
\citep[in the age range between 20 Myr and 300 Myr, see][]{renzini}, 
and were much more massive in the past \citep[see e.g.][]{dercole,vesperini10,conroy}
to be able to retain the stellar ejecta,  
younger and less massive clusters
did not undergo self-enrichment processes and their present-day mass should be very similar 
to their initial mass \citep[see the discussion in][]{m11}.
Thus, the observational evidences presented here confirm 
the theoretical predictions that clusters with initial mass below  $10^{5} M_{\odot}$ should 
be chemically homogeneous \citep[see][]{bland}.
It is worth to notice that stellar clusters with ages of $\sim$100 Myr and 
masses of $\sim$1-5 $10^{4} M_{\odot}$ are lacking in our Galaxy, the 
open clusters in the Milky Way being at least 1 order of magnitude less massive than 
the coeval LMC globulars. Hence, the study of such young LMC clusters appears to 
be particularly illuminating to understand the early evolution of the globulars.

Concerning the final fate of the NGC~2136/NGC~2137 system, 
two possible scenario can be prospected: 
(i)~the two clusters will finally merge under the 
action of the dynamical friction; or (ii)~the 
mutual tidal forces will disrupt the less massive system, dispersing its stellar 
content (and probably leaving a weak stellar stream around the survived cluster).

The dynamical friction timescale, hence the time for the secondary cluster to 
spiral into the main cluster,
can be estimated with Eq.7-26 of \citet{binney87}, assuming 
the present-day conditions. 
We derive a merging time-scale 
of $\sim$38 Myr, comparable with the orbital period. 
However, this scenario is reliable
only if the secondary cluster crosses the main cluster: 
\citet{mclau} derive for NGC~2136 
a tidal radius of 30.9 pc by adopting a King model, 
larger than the projected distance between 
the two clusters ($\sim 20$ pc).  Hence, we can consider realistic 
the occurrence of dynamical friction between 
the two objects.
Alternatively, NGC~2137 will be destroyed by the tidal field of NGC~2136, in a 
time-scale of $\sim$2 
Gyr, estimated by adopting Eq. 1 by \citet{gieles}.
Due to the uncertainty in the mass values and in the 3-D distance,
these calculations should be considered as a first-order  estimation
of the  timescales. 

The perspective of a merging is quite interesting in light of the 
formation history of the LMC GCs.
Numerical simulations by \citet{makino91}, \citet{deoliv} and \citet{bekki} 
predict that the final merger of a binary cluster would be indistinguishable 
from a genuine single-population GC, but with high values of ellipticity 
($\epsilon$=~0.25-0.35), very similar to those observed in the LMC GCs 
\citep[see e.g.][]{geisler80,m07}. In this framework, we cannot exclude that 
a fraction of the present-day single LMC clusters were originated by the
merging of twin objects (in terms of kinematics, age and chemical composition).

\section{Conclusions}

The analysis of the kinematical and chemical properties of the pair
NGC~2136/NGC~2137 presented in this paper demonstrate that the two clusters
share not only the same age \citep{hilker,dirsch} 
but also the same chemical and  kinematic {\sl DNA}, 
unequivocally ensuring their common origin.
{\sl This is the first firm validation of the true binary nature of a LMC cluster pair}: 
the previous hints were in fact only based on their projected distances and photometric properties.
From the obtained results we can also draw a general scenario for the formation and 
evolution of this binary system. Summarizing: (i)~the two clusters formed from the 
fragmentation of the same molecular cloud; 
(ii)~the chemical composition of the clusters is about the same, without 
hints of self-enrichment or mutual chemical pollution; (iii)~the present-day 
orbital parameters suggest that the system will merge due to the 
dynamical friction in a timescale comparable with its orbital period.

An important point to tackle is the different frequency of 
candidate binary clusters in the LMC and in the Milky Way.
Theoretical models by \citet{fuji} and \citet{bekki} demonstrate that binary 
stellar clusters can be formed during high velocity cloud collisions. 
Concerning the LMC, the rate of cloud collisions is mainly triggered by the mutual 
tidal interaction between the LMC and the SMC, and their close encounters 
\citep[the last one occuring $\sim$200 Myr ago][]{bekki05}
On the other hand, the Galactic disk is less disturbed by the 
near tidal fields, at variance to the LMC
(that suffers of the effects due to the interaction with the Galaxy and SMC fields),
and the rate of cloud-cloud collisions is less efficient, as demonstrated by the dearth 
of binary systems among the Milky Way open clusters.
Thus, the occurrence of binary clusters is intimately linked to the star formation 
history of their parent galaxy and the interactions of the latter with near tidal fields.

Our findings indicate that other candidate cluster pairs in the LMC 
could be probed to be binary systems through the analysis of high-resolution spectra. 
Direct kinematical and 
chemical measurements of such systems (with both similar and different component clusters ages) 
are mandatory to assess the origin of these systems and enlighten on the cluster 
formation history in the Magellanic Clouds.\\

\acknowledgements  
We warmly thank the anonymous referee for his/her suggestions in improving the paper.
This research is part of the project COSMIC-LAB funded by the European Research Council 
(under contract ERC-2010-AdG-267675).

{}

\begin{figure}  
\epsscale{0.8}   
\plotone{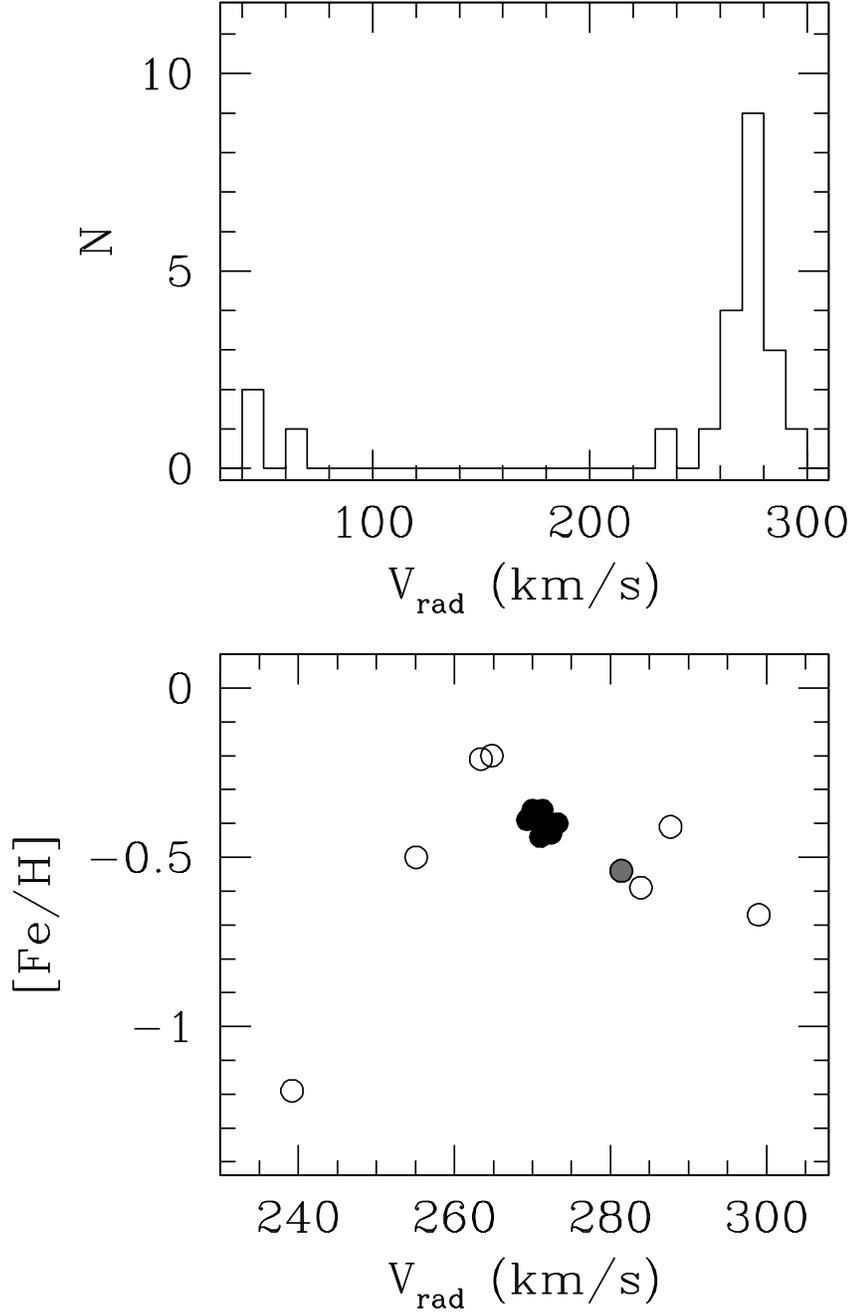}
\caption{{\it Upper panel:}  the radial velocities distribution for the observed targets in the 
inner 3 arcmin from the center of NGC~2136 (see Fig.~\ref{map}). 
{\it Lower panel:} the distribution of the stars with $V_{r}>$230 km/s in the $V_{r}$-[Fe/H] plane: 
black circles are the stars of the cluster pair, the empty circles are the stars 
belonging to the LMC field and the grey circle is the giant located in the 
possible third component of the system suggested by \citep{hilker}.
}
\label{vrad}
\end{figure}

\begin{figure} 
\plotone{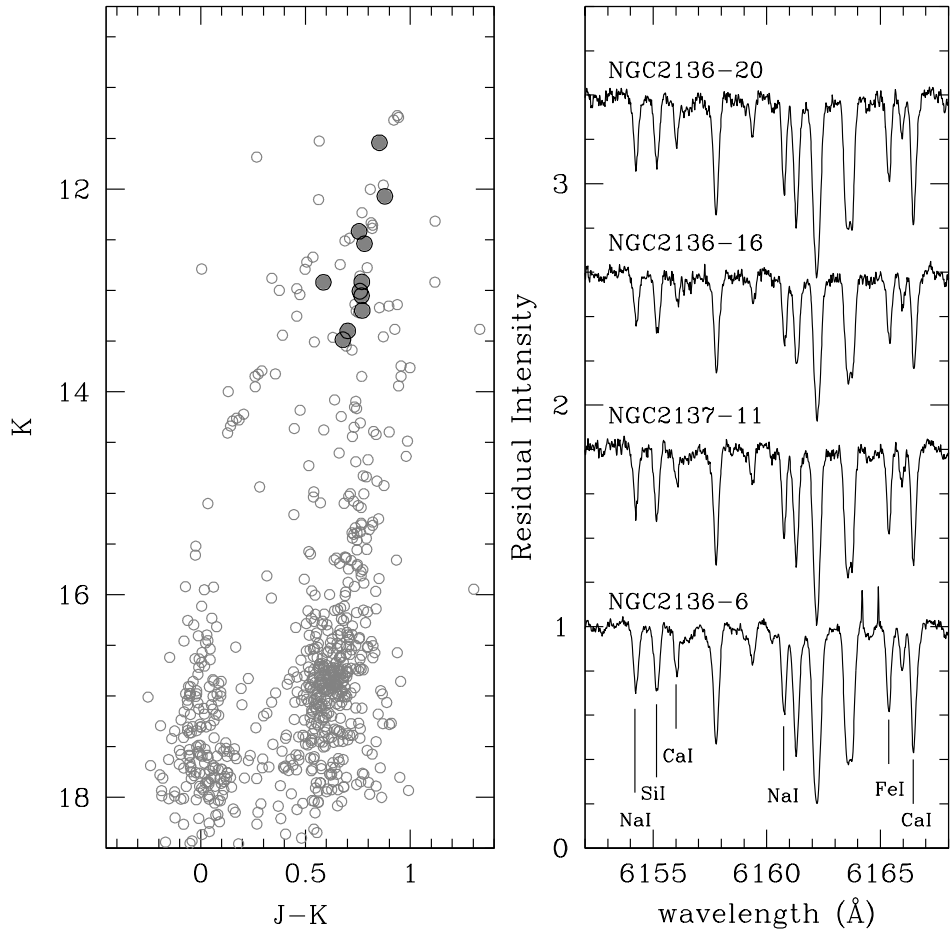}
\caption{{\sl Left panel}: SofI color-magnitude diagram of the field around NGC~2136 and NGC~2137; large grey 
points mark the spectroscopic targets. {\sl Right panel}: UVES spectra of four stars. A few reference 
lines are marked.}
\label{cmd}
\end{figure}

\begin{landscape}
\begin{deluxetable}{lccccccccc}
\tablecolumns{10} 
\tiny
\tablewidth{0pc}  
\tablecaption{}
\tablehead{ 
\colhead{Star} &   RA & 
Dec & J & K & $V_{r}$ & $T_{eff}$ & log~g & $v_t$ &
[Fe/H] \\
  &   (J2000)  &  (J2000)  &  & &(km/s) & (K) & & (km/s)  & (dex)  }
\startdata 
\hline
 NGC~2136-6   &  88.2319114   &  -69.4903000     &  12.39   & 11.54  &  271.0$\pm$0.5	& 4100  &  0.90  & 2.10  & --0.44$\pm$0.07       \\    
 NGC~2136-16  &  88.2304271   &  -69.4941896     &  13.17   & 12.42  &  272.4$\pm$0.7	& 4400  &  1.40  & 1.50  & --0.43$\pm$0.07       \\    
 NGC~2136-20  &  88.2396754   &  -69.4952318     &  13.32   & 12.54  &  271.1$\pm$0.9	& 4150  &  1.20  & 1.90  & --0.39$\pm$0.07       \\    
 NGC~2136-31  &  88.2214002   &  -69.4928704     &  13.68   & 12.92  &  271.3$\pm$0.5	& 4400  &  1.50  & 1.80  & --0.36$\pm$0.10       \\    
 NGC~2136-34  &  88.2469143   &  -69.4870399     &  13.77   & 13.01  &  273.2$\pm$0.5	& 4400  &  1.50  & 1.70  & --0.40$\pm$0.09       \\    
 NGC~2136-35  &  88.2270944   &  -69.4809438     &  13.82   & 13.05  &  271.7$\pm$0.8	& 4400  &  1.60  & 1.80  & --0.40$\pm$0.05     \\	 
 NGC~2136-46  &  88.2052808   &  -69.4879732     &  14.04   & 12.92  &  270.0$\pm$0.6	& 4550  &  1.80  & 1.50  & --0.36$\pm$0.07     \\	 
\hline 
 NGC~2137-11  &  88.3056357   &  -69.4811137     &  12.95   & 12.07  &  271.0$\pm$0.4	& 4350  &  1.50  & 1.80  & --0.38$\pm$0.09       \\    
 NGC~2137-28  &  88.3056203   &  -69.4855993     &  13.51   & 12.92  &  271.1$\pm$0.5	& 5000  &  1.10  & 2.50  & --0.40$\pm$0.05       \\    
 NGC~2137-41  &  88.2970989   &  -69.4820216     &  13.97   & 13.20  &  271.0$\pm$0.9	& 4450  &  1.80  & 1.50  & --0.40$\pm$0.10       \\    
 NGC~2137-51  &  88.2714259   &  -69.4785592     &  14.16   & 13.49  &  269.3$\pm$0.6	& 4750  &  1.90  & 1.60  & --0.39$\pm$0.08       \\    
\hline
\hline
\colhead{Star} & [O/Fe]  &  [Na/Fe]  &  [Mg/Fe]  &  [Al/Fe]  &  [Si/Fe]  &  [Ca/Fe]  &  [Ti/Fe]  &  [Ni/Fe] & \\
 & (dex)  &   (dex)   &  (dex)    & (dex) & (dex) &(dex) & (dex) & (dex) & \\ 
\hline 
  NGC~2136-6    &   0.04$\pm$0.05	&    --0.23$\pm$0.06  &   0.06$\pm$0.08    &--0.03$\pm$0.06    &    0.12$\pm$0.10   &--0.07$\pm$0.08	&   --0.08$\pm$0.10   &  --0.16$\pm$0.05   &	       \\ 
  NGC~2136-16   & --0.09$\pm$0.07	&    --0.24$\pm$0.06  & --0.02$\pm$0.07    &--0.07$\pm$0.07    &    0.05$\pm$0.12   &--0.04$\pm$0.08	&   --0.08$\pm$0.07   &  --0.18$\pm$0.05   &	       \\ 
  NGC~2136-20   &   0.00$\pm$0.04	&    --0.23$\pm$0.08  &   0.02$\pm$0.10    &--0.12$\pm$0.08    &    0.16$\pm$0.11   &--0.07$\pm$0.10	&   --0.11$\pm$0.08   &  --0.18$\pm$0.05   &	       \\ 
  NGC~2136-31   &   0.02$\pm$0.06	&    --0.22$\pm$0.05  &   0.01$\pm$0.08    &--0.10$\pm$0.06    &    0.09$\pm$0.12   &  0.00$\pm$0.07	&   --0.09$\pm$0.12   &  --0.20$\pm$0.05   &	       \\ 
  NGC~2136-34   & --0.03$\pm$0.06	&    --0.23$\pm$0.08  &   0.04$\pm$0.11    &--0.08$\pm$0.06    &    0.13$\pm$0.09   &  0.01$\pm$0.10	&   --0.09$\pm$0.10   &  --0.13$\pm$0.06   &	       \\ 
  NGC~2136-35   & --0.01$\pm$0.08	&    --0.21$\pm$0.06  & --0.05$\pm$0.07    &   ---             &    0.04$\pm$0.12   &  0.04$\pm$0.10    &   --0.13$\pm$0.09   &  --0.09$\pm$0.06   &  \\ 
  NGC~2136-46   &   0.02$\pm$0.07	&    --0.27$\pm$0.05  & --0.08$\pm$0.07    &   ---             &    0.04$\pm$0.11   &--0.09$\pm$0.12    &   --0.14$\pm$0.12   &  --0.16$\pm$0.07   &   \\ 
\hline  
  NGC~2137-11	& --0.03$\pm$0.06	&    --0.18$\pm$0.06  &   0.02$\pm$0.12    &--0.08$\pm$0.06    &    0.12$\pm$0.10   &--0.03$\pm$0.11    &   --0.08$\pm$0.12   &  --0.15$\pm$0.05   &   \\ 
  NGC~2137-28	& --0.07$\pm$0.08	&    --0.22$\pm$0.07  &   0.05$\pm$0.11    &   ---             &    0.01$\pm$0.10   &--0.08$\pm$0.07    &   --0.18$\pm$0.14   &  --0.20$\pm$0.05   &	    \\ 
  NGC~2137-41	&   0.03$\pm$0.09	&    --0.36$\pm$0.07  &   0.05$\pm$0.09    &   ---             &    0.21$\pm$0.12   &  0.01$\pm$0.12    &   --0.14$\pm$0.10   &  --0.17$\pm$0.07   &  \\ 
  NGC~2137-51	&   0.00$\pm$0.07	&    --0.24$\pm$0.08  &   0.01$\pm$0.10    &   ---             &    0.08$\pm$0.08   &  0.02$\pm$0.10    &   --0.02$\pm$0.12   &  --0.15$\pm$0.06   &  \\ 
\hline
\enddata 
\tablecomments{$~~~~~$Main information of the member clusters stars. Coordinates and magnitudes are 
from \citet{m06}.The uncertainties in $V_{r}$ include the internal error and the uncertainty 
in the wavelength calibration zero-point.  The abundance uncertainties  include the internal error and 
that due to the atmospheric parameters.}
\end{deluxetable}
\end{landscape}

\begin{deluxetable}{ccccc}
\tablecolumns{5} 
\tablewidth{0pc}  
\tablecaption{}
\tablehead{ 
 Ratio &  NGC~2136 &    &  NGC~2137&    \\
    &    mean   &  $\sigma$   & mean  & $\sigma$  }
\startdata 
\hline
 \colhead{[Fe/H]}   &  -0.40     &  0.03    &   -0.39  &   0.01     \\    
 \colhead{[O/Fe]}   &  -0.01     &  0.04    &	-0.02  &   0.04    \\	 
 \colhead{[Na/Fe]}  &  -0.24     &  0.02    &   -0.25  &   0.08   \\      
 \colhead{[Mg/Fe]}  &  +0.00     &  0.05    &   +0.03  &   0.02	 \\    
 \colhead{[Al/Fe]}  &  -0.08     &  0.03    &   -0.08  &    ---	 \\    
 \colhead{[Si/Fe]}  &  +0.14     &  0.05    &   +0.15  &   0.08	 \\     
 \colhead{[Ca/Fe]}  &  -0.03     &  0.05    &   -0.02  &   0.04	 \\    
 \colhead{[Ti/Fe]}  &  -0.10     &  0.02    &   -0.10  &   0.07	 \\     
 \colhead{[Ni/Fe]}  &  -0.16     &  0.04    &   -0.17  &   0.02	 \\	
\hline
\enddata 
\tablecomments{Solar reference abundances are from \citet{gs}, 
with the exception of the oxygen \citep{caffau}.}
\end{deluxetable}

\end{document}